\documentclass[10pt,twoside,twocolumn,american,english,aps,nofootinbib,preprintnumbers,superscriptaddress]{revtex4}
\usepackage[T1]{fontenc}
\usepackage[latin9]{inputenc}
\usepackage{color}
\usepackage{babel}
\usepackage{mathrsfs}
\usepackage{amsmath}
\usepackage{amssymb}
\usepackage[unicode=true,pdfusetitle,
 bookmarks=true,bookmarksnumbered=false,bookmarksopen=false,
 breaklinks=true,pdfborder={0 0 0},pdfborderstyle={},backref=false,colorlinks=true]
 {hyperref}

\makeatletter
\@ifundefined{textcolor}{}
{%
 \definecolor{BLACK}{gray}{0}
 \definecolor{WHITE}{gray}{1}
 \definecolor{RED}{rgb}{1,0,0}
 \definecolor{GREEN}{rgb}{0,1,0}
 \definecolor{BLUE}{rgb}{0,0,1}
 \definecolor{CYAN}{cmyk}{1,0,0,0}
 \definecolor{MAGENTA}{cmyk}{0,1,0,0}
 \definecolor{YELLOW}{cmyk}{0,0,1,0}
}

\usepackage{hyperref}
\hypersetup{
    colorlinks,%
    citecolor=blue,%
    filecolor=blue,%
    linkcolor=blue,%
    urlcolor=blue
}
\usepackage{mathtools}

\makeatother

\begin{document}
\title{Losing the trace to find dynamical Newton or Planck constants}
\author{Pavel Jirou\v{s}ek}
\email{jirousek@fzu.cz}

\affiliation{CEICO--Central European Institute for Cosmology and Fundamental Physics,
\\
FZU--Institute of Physics of the Czech Academy of Sciences, \\
Na Slovance 1999/2, 18221 Prague 8, Czech Republic }
\affiliation{Institute of Theoretical Physics, Faculty of Mathematics and Physics,
Charles University, \\
V Hole\v{s}ovi\v{c}k\'ach 2, 180 00 Prague 8, Czech Republic}
\author{Keigo Shimada}
\email{shimada.k.ah@m.titech.ac.jp}

\affiliation{Department of Physics, Tokyo Institute of Technology, Tokyo 152-8551,
Japan}
\author{Alexander Vikman}
\email{vikman@fzu.cz}

\affiliation{CEICO--Central European Institute for Cosmology and Fundamental Physics,
\\
FZU--Institute of Physics of the Czech Academy of Sciences, \\
Na Slovance 1999/2, 18221 Prague 8, Czech Republic }
\author{Masahide Yamaguchi}
\email{gucci@phys.titech.ac.jp}

\affiliation{Department of Physics, Tokyo Institute of Technology, Tokyo 152-8551,
Japan}
\date{\today}
\begin{abstract}
We show that promoting the trace part of the Einstein equations to
a trivial identity results in the Newton constant being an integration
constant. Thus, in this formulation the Newton constant is a global
dynamical degree of freedom which is also a subject to quantization
and quantum fluctuations. This is similar to what happens to the cosmological
constant in the unimodular gravity where the trace part of the Einstein
equations is lost in a different way. We introduce a constrained variational
formulation of these modified Einstein equations. Then, drawing on
analogies with the Henneaux-Teitelboim action for unimodular gravity,
we construct different general-covariant actions resulting in these
dynamics. The inverse of dynamical Newton constant is canonically
conjugated to the Ricci scalar integrated over spacetime. Surprisingly,
instead of the dynamical Newton constant one can formulate an equivalent
theory with a dynamical Planck constant. Finally, we show that an
axion-like field can play a role of the gravitational Newton constant
or even of the quantum Planck constant. 
\end{abstract}
\maketitle

\section{Introduction and Our Main Idea }

The origin of the cosmological constant and its relation to the vacuum
fluctuations of quantum fields persists to be one of the main puzzles
of contemporary physics, see e.g. \citep{Zeldovich:1968ehl,Weinberg:1988cp,Martin:2012bt,Padilla:2015aaa}.
However, since Einstein's seminal paper \citep{Einstein:1919gv} it
is well known that trace-free equations
\begin{equation}
G_{\mu\nu}-\frac{1}{4}g_{\mu\nu}G=8\pi G_{N}\left(T_{\mu\nu}-\frac{1}{4}g_{\mu\nu}T\right)\,,\label{eq:Unimod}
\end{equation}
result in a cosmological constant (CC), $\Lambda$, as a constant
of integration, or \emph{a global} degree of freedom, see e.g. \citep{Anderson:1971pn,Weinberg:1988cp,Unruh:1988in,Unruh:1989db,Finkelstein:2000pg,Ellis:2010uc,Ellis:2013eqs}.
For all other purposes, the trace-free equations above are entirely
classically equivalent to the usual General Relativity (GR) equations
with $\Lambda$
\begin{equation}
G_{\mu\nu}=8\pi G_{N}T_{\mu\nu}+\Lambda g_{\mu\nu}\,.\label{eq:Usual_Einstein}
\end{equation}

Crucially, the trace-free Einstein equations (\ref{eq:Unimod}) are
invariant under vacuum shifts of the total energy-momentum tensor
(EMT)
\begin{equation}
T_{\mu\nu}\rightarrow T_{\mu\nu}+c\,g_{\mu\nu}\,,\label{eq:vacuum_shifts}
\end{equation}
with $c=c\left(x^{\mu}\right)$, and in particular with $c=const$,
which is not the case for the usual Einstein equations (\ref{eq:Usual_Einstein}),
with or without $\Lambda$.

Nothing illustrates better the currently unexplainable small value
of $\Lambda$, as a comparison with $G_{N}$ --- the only other dimensionful\footnote{We use $\left(+,-,-,-\right)$ signature convention and units where
$c=1$, while in most parts of the text $\hbar=1$ and we restore
$\hbar$ only to stress the quantum nature of a formula.} constant in (\ref{eq:Usual_Einstein}):
\begin{equation}
G_{N}\Lambda\sim10^{-122}\,.\label{eq:120}
\end{equation}
The usual question to ask regarding these 122 orders of magnitude
is: why is $\Lambda$ so vanishingly small, but still not entirely
zero? Perhaps one should invert the question and ask instead: why
is $G_{N}$ that tiny? Both these questions require an ability to
select values on these constants from some ensemble. Thus, it makes
sense to look for simple theories where $G_{N}$ would appear as an
integration constant or a global degree of freedom. This is a ``landscape''
of poor people. 

With this motivation in mind, in this paper we propose a novel way
to lose the trace part of the Einstein equations. Indeed, instead
of (\ref{eq:Unimod}) one can write \emph{normalized} or \emph{trace-trivial}
equations 
\begin{equation}
\frac{G_{\mu\nu}}{G}=\frac{T_{\mu\nu}}{T}\,.\label{eq:normalized-2}
\end{equation}
These equations are \emph{scale-free }and invariant with respect to
rescaling\footnote{It is worth noting that transformations (\ref{eq:vacuum_shifts})
and (\ref{eq:rescale}) could be performed on the Einstein tensor
instead of the EMT, see e.g. \citep{Alexander:2018tyf}. However,
it is not clear how to realize these transformations as transformations
of the metric. Moreover, the equations are completely blind to independent
transformations of both tensors. } 
\begin{equation}
T_{\mu\nu}\rightarrow c\,T_{\mu\nu}\,,\label{eq:rescale}
\end{equation}
by a constant, or even an arbitrary function $c\left(x^{\mu}\right)$.
As we show in this paper, equations (\ref{eq:normalized-2}) are again
equivalent to the usual GR, however now $G_{N}$ becomes a constant
of integration. 

Following Einstein \citep{Einstein:1919gv} on the cosmological constant,
one can either conclude that \textquotedblleft ... the new formulation
has this great advantage, that the quantity appears in the fundamental
equations as a constant of integration, and no longer as a universal
constant peculiar to the fundamental law\textquotedblright{} or rather
that ``Since the world exists as a single specimen, it is essentially
the same whether a constant is given the form of one belonging to
the natural laws or the form of an \textquoteleft integration constant\textquoteright \textquotedblright ,
as in \citep{Einstein:1918c}\footnote{For history of the cosmological constant see e.g. \citep{ORaifeartaigh:2017yby}}.
Crucially, quantum cosmology and chaotic self-reproducing inflation
\citep{Vilenkin:1983xq,Linde:1986fd} allow for a multiverse so that
the world would be not a ``single specimen'', as thought by Einstein.
Thus, one can speculate that the value of the Newton constant may
be a remnant from the early universe quantum gravity era, similarly
to what could happen to the cosmological constant, as discussed in
e.g. \citep{Linde:1984ir,Linde:2015edk}. 

\section{Recap of unimodular gravity}

Our proposal shares some structural similarities with the unimodular
gravity. Hence, it is worthwhile to recall its basic properties here.
As described in e.g. \citep{Anderson:1971pn,Weinberg:1988cp,Unruh:1988in,Unruh:1989db,Finkelstein:2000pg,Ellis:2010uc,Ellis:2013eqs},
one can apply covariant derivative $\nabla_{\mu}$ to both sides of
(\ref{eq:Unimod}) and by virtue of the Bianchi identity, $\nabla_{\mu}G^{\mu\nu}=0$,
along with the \emph{assumed} conservation of EMT, $\nabla_{\mu}T^{\mu\nu}=0$,
one obtains\footnote{In some formulations of gravity, the Bianchi identity may not hold
which results in non-vanishing right hand side of this differential
consequence. The implications of this have been recently explored
in \citep{Alexander:2019ctv,Alexander:2019wne,Magueijo:2019vmk}.}
\begin{equation}
\partial_{\mu}\left(G-\varkappa T\right)=0\,,\label{eq:Consequence}
\end{equation}
where we introduced notation 
\begin{equation}
\varkappa=8\pi G_{N}\,,\label{eq:kappa}
\end{equation}
to make later expressions shorter. In this way 
\begin{equation}
G-\varkappa T=4\Lambda\,,\label{eq:Lambda}
\end{equation}
where $\Lambda$ is a constant of integration. Substitution of this
differential consequence back into the trace-free Einstein equations
(\ref{eq:Unimod}) yields normal Einstein's General Relativity (GR)
with the cosmological constant $\Lambda$. Thus, loosing the trace-part
of the standard Einstein equations in this way results in promoting
(or demoting) the cosmological constant (CC) from the fundamental
constants of nature, parameters or coupling constants to a global
dynamical degree of freedom. This modification of GR is known under
the name ``unimodular gravity''. 

One can obtain the traceless Einstein equations (\ref{eq:Unimod}),
provided one varies the Einstein-Hilbert action under \emph{constrained}
or \emph{restricted} variation of the contravariant metric, $\bar{\delta}g^{\mu\nu}$,
satisfying 
\begin{equation}
g_{\mu\nu}\bar{\delta}g^{\mu\nu}=0\,.\label{eq:transversality}
\end{equation}
This variational constraint implies that only \emph{transverse} variations
$\bar{\delta}g^{\mu\nu}$ are allowed. This general-covariant restriction
(\ref{eq:transversality}) of the variations $\delta g^{\mu\nu}$
can be obtained from an apparently non-covariant ``unimodular''
constraint 
\begin{equation}
\sqrt{-g}\equiv\sqrt{-\text{det}g_{\mu\nu}}=\sigma\left(x\right)\,,\label{eq:unimodular_condition}
\end{equation}
where $\sigma\left(x^{\mu}\right)$ is an unspecified \emph{function}
which is often taken to be just unity: 
\begin{equation}
\sqrt{-g}=1\,.\label{eq:unimod_bad}
\end{equation}
However, the later choice manifestly breaks general covariance and
would select preferred coordinate systems. Moreover, this selection
clearly excludes writing most important solutions of GR like the Schwarzschild
and Friedmann metrics in natural coordinates which have simple physical
interpretation and which can definitely be formed by test bodies and
clocks. Clearly, for every metric, $g_{\mu\nu}$, there is enough
gauge freedom to find a coordinate system to satisfy the ``unimodular''
constraint (\ref{eq:unimod_bad}), but it is especially easy if one
does not specify the function $\sigma\left(x\right)$ at all. The
irrelevant choice of this function corresponds to a gauge freedom
responsible for the diffeomorphism invariance. In the manifestly general-covariant
formulation of the unimodular gravity by Henneaux-Teitelboim \citep{Henneaux:1989zc}\footnote{For other diff invariant formulations of the unimodular gravity, see
e.g. \citep{Kuchar:1991xd,Jirousek:2018ago,Hammer:2020dqp}} the meaning of the ``unimodular'' constraint is reversed: it is
used not to find $\sqrt{-g}$ for a given $\sigma\left(x\right)$,
but to find instead a dynamical variable composing $\sigma$ for a
given $\sqrt{-g}$. Most importantly, restriction on variations (\ref{eq:transversality})
is covariant and so are the resulting traceless equations of motion
(\ref{eq:Unimod}). Thus, it is rather counterproductive to take $\sqrt{-g}=1$
and break their general covariance by this unnecessary assumption,
even though this choice makes the action polynomial \citep{vanderBij:1981ym}
and may simplify some calculations. 

To quickly see that (\ref{eq:transversality}) results in (\ref{eq:Unimod})
one writes first variation form of the Einstein equations for \emph{general}
variations of the metric which is also satisfied by the restricted
variations $\bar{\delta}g^{\mu\nu}$
\begin{equation}
\left(G_{\mu\nu}-\varkappa T_{\mu\nu}\right)\bar{\delta}g^{\mu\nu}=0\,.\label{eq:unrestricted_variation}
\end{equation}

However, due to the transversality constraint (\ref{eq:transversality})
only tensor components not ``proportional'' to $g_{\mu\nu}$ had
to be equal to zero now. This results in appearance of the Lagrange
multiplier $\lambda\left(x^{\mu}\right)$ responsible for the ``reaction
force'' of the constraint 
\begin{equation}
G_{\mu\nu}-\varkappa T_{\mu\nu}=\lambda\left(x\right)g_{\mu\nu}\,.\label{eq:1Kind_Lagrange_Unimod}
\end{equation}
This is an analog of the Lagrange's equations of the first kind in
analytical mechanics. One can exclude unspecified Lagrange multiplier
by taking the trace of (\ref{eq:1Kind_Lagrange_Unimod}), so that
$\lambda\left(x\right)=\left(G-\varkappa T\right)/4,$ and then substituting
this expression back to (\ref{eq:1Kind_Lagrange_Unimod}) one obtains
trace-free Einstein equations (\ref{eq:Unimod}). Of course, we could
apply covariant derivative and use Bianchi identity along with the
covariant conservation of EMT directly in (\ref{eq:1Kind_Lagrange_Unimod})
to find that $\lambda\left(x\right)=const$. 

It is well known how to write an action for the trace-free Einstein
equations (\ref{eq:Unimod}). The arguably simplest covariant action
was introduced by Henneaux and Teitelboim in \citep{Henneaux:1989zc}
\begin{equation}
S\left[g,W,\Lambda\right]=\frac{1}{\varkappa}\int d^{4}x\sqrt{-g}\left[-\frac{R}{2}+\Lambda\left(\nabla_{\mu}W^{\mu}-1\right)\right]\,.\label{eq:vector_for_CC}
\end{equation}
The second term in the brackets has a form of the Hamiltonian (or
first order) action. Furthermore, variation with respect to the spatial
components $W^{i}$ yields $\partial_{i}\Lambda=0$, along every Cauchy
hypersurface $\Sigma$. Hence, for every imposed foliation, $\Lambda$
is space independent - is a global quantity. Using this constraint
and following Faddeev-Jackiw \citep{Faddeev:1988qp,Jackiw:1993in},
one concludes ``without tears'' that $\Lambda/\varkappa$ is a canonical
momentum conjugated to the cosmic time 
\begin{equation}
\mathscr{T}\left(t\right)=\int_{\Sigma}d^{3}\mathbf{x}\sqrt{-g}\,W^{t}\left(t,\mathbf{x}\right)\,.\label{eq:cosmic_time}
\end{equation}
This variable measures four-volume of space-time between Cauchy hypersurfaces
$\Sigma_{2}$ and $\Sigma_{1}$ as 
\begin{equation}
\mathscr{T}\left(t_{2}\right)-\mathscr{T}\left(t_{1}\right)=\int_{\mathscr{V}}d^{4}x\sqrt{-g}\,.\label{eq:4volume}
\end{equation}
One can think about $\mathscr{T}$ as a charge which is not conserved,
but rather continuously produced by a unit source due to the constraint
\begin{equation}
\nabla_{\mu}W^{\mu}=1\,.\label{eq:constraint_W}
\end{equation}
As $\Lambda/\varkappa$ and $\mathscr{T}$ are canonically conjugated,
one can apply Heisenberg uncertainty relation 
\begin{equation}
\delta\Lambda\times\delta\int_{\mathscr{V}}d^{4}x\sqrt{-g}\geq4\pi\,\ell_{Pl}^{2}\,,\label{eq:uncertainty_volume}
\end{equation}
where $\ell_{Pl}=\sqrt{\hbar G_{N}}$ is the Planck length. The presence
of these quantum fluctuations is the main difference of ``unimodular''
gravity from the usual GR. These fluctuations may be relevant close
to singularities. Except of this phenomenon, perturbative ``unimodular''
gravity is equivalent to GR also in quantum realm, for recent discussions
see e.g. \citep{Alvarez:2005iy,Fiol:2008vk,Smolin:2009ti,Eichhorn:2013xr,Saltas:2014cta,Padilla:2014yea,Bufalo:2015wda,Alvarez:2015sba,Alvarez:2015pla,Percacci:2017fsy,Ardon:2017atk,Herrero-Valea:2018ilg,Herrero-Valea:2020xaq,deBrito:2020rwu}.

The constraint part of the action (\ref{eq:vector_for_CC}) can be
written as 
\begin{equation}
\int d^{4}x\,\Lambda\left[\partial_{\mu}\left(\sqrt{-g}W^{\mu}\right)-\sqrt{-g}\,\right]\,,\label{eq:First_step_non_covariant}
\end{equation}
so that $\partial_{\mu}\left(\sqrt{-g}W^{\mu}\right)$ plays the role
of the unspecified function $\sigma\left(x\right)$ in (\ref{eq:unimodular_condition})
or the measure\footnote{Interestingly, this can be considered as a particular example of theories
with two different measures, see e.g. \citep{Guendelman:1996qy,Guendelman:1999qt,Guendelman:1999tb}. } from \citep{Anderson:1971pn}. Though, now the ``unimodular'' constraint
(\ref{eq:unimodular_condition}) is used to find $W^{\mu}$ without
imposing any restrictions on the metric. 

On the other hand, action (\ref{eq:vector_for_CC}) possesses a gauge
redundancy 
\begin{equation}
W^{\mu}\rightarrow W^{\mu}+\epsilon^{\mu}\,,\quad\text{where}\quad\nabla_{\mu}\epsilon^{\mu}=0\,.\label{eq:gauge_redundancy_W}
\end{equation}
 For every $W^{\mu}$ and every choice of coordinates $\left(t',x^{i}\right)$
one can find such $\epsilon^{\mu}$ that the spatial components vanish
$W^{\mu}=(W^{t'},0)$. Further, one can introduce new coordinates
$\left(t,x^{i}\right)$ with $t\left(t',x^{i}\right)=\sqrt{-g'}W^{t'}$,
so that 
\begin{equation}
W^{t}=\frac{\partial t}{\partial t'}W^{t'}=\frac{\partial t}{\partial t'}\frac{t}{\sqrt{-g'}}\,,\label{eq:W_transform}
\end{equation}
which through the transformation of measure $\sqrt{-g}\,\partial t/\partial t'=\sqrt{-g'}$
implies that 
\begin{equation}
W^{\mu}=\delta_{t}^{\mu}\frac{t}{\sqrt{-g}}\,.\label{eq:fixed_W}
\end{equation}
If we fix this gauge and these coordinates \emph{before} performing
the variation (what is usually not a correct way to proceed) we obtain
the often used ``unimodular'' constraint\footnote{In terms of the ADM variables \citep{Arnowitt:1962hi,Poisson} the
``unimodular'' constraint relates the lapse $N$ to the determinant
of the spatial metric $\gamma$ as $N=\gamma^{-1/2}$. This relation
can be further extended \citep{Barvinsky:2017pmm} to a general function
$N=f\left(\gamma\right)$, with rather peculiar consequences, e.g.
\citep{Barvinsky:2019agh,Barvinsky:2020sxl,Barvinsky:2019qzx}.} 
\begin{equation}
\int d^{4}x\,\Lambda\left[1-\sqrt{-g}\,\right]\,,\label{eq:fixed_W_constraint}
\end{equation}
which is a usual starting point for a non-covariant formulation of
the unimodular gravity \citep{Buchmuller:1988wx,Buchmuller:1988yn,vanderBij:1981ym,Alvarez:2006uu}.
Sometimes it is claimed that this constraint (\ref{eq:fixed_W_constraint})
can be satisfied just by a proper choice of the coordinates. However,
it is worth emphasizing that on the way to (\ref{eq:fixed_W_constraint})
we have already used the diffeomorphism invariance once to enforce
the unity of the density $\partial_{\mu}\left(\sqrt{-g}W^{\mu}\right)$.
If we could use the diffeomorphism again to enforce the unity of $\sqrt{-g}$,
this would imply that just by a diffeomorphism one could enforce the
unity of the scalar $\nabla_{\mu}W^{\mu}$ which is clearly not possible.
Thus, constraint (\ref{eq:fixed_W_constraint}) is not just a mere
gauge choice, but does change the dynamics by introducing a global
degree of freedom. 

As showed in \citep{Hammer:2020dqp}, the vector field $W^{\mu}$
can be exchanged with a more convenient Chern-Simons current of a
(abelian or non-abelian) gauge field $A_{\mu}$ so that 
\begin{equation}
S\left[g,A,\Lambda\right]=\frac{1}{\varkappa}\int d^{4}x\sqrt{-g}\left[-\frac{R}{2}+\Lambda\left(F_{\alpha\beta}F^{\star\alpha\beta}-1\right)\right]\,,\label{eq:Chern_Simons_CC}
\end{equation}
where $F_{\mu\nu}=D_{\mu}A_{\nu}-D_{\nu}A_{\mu}$ is the gauge field
strength or curvature of the covariant derivative $D_{\mu}=\nabla_{\mu}+iqA_{\mu}$,
while the Hodge dual is defined as 
\begin{equation}
F^{\star\alpha\beta}=\frac{1}{2}\cdot\frac{\epsilon^{\alpha\beta\mu\nu}}{\sqrt{-g}}\cdot F_{\mu\nu}\equiv\frac{1}{2}\cdot E^{\alpha\beta\mu\nu}\cdot F_{\mu\nu}\,.\label{eq:dual}
\end{equation}
Moreover, one could avoid the Lagrange multiplier and formulate unimodular
gravity as a higher-derivative and Weyl-invariant theory of a vector
field \citep{Jirousek:2018ago} or of a gauge field \citep{Hammer:2020dqp},
see also \citep{Kimpton:2012rv}. One can also describe this dynamics
by a covariant action containing a three-form as in e.g. \citep{Aurilia:1980xj,Henneaux:1984ji},
but we will not expand on that. 

Finally, we would like to mention that it is the absence of the usual
kinetic term $F_{\alpha\beta}F^{\alpha\beta}$ which forces $\Lambda$
to be constant. Hence, it is easy to further extend this action to
resemble the one of the usual axion, c.f. \citep{Wilczek:1983as}
and see \citep{Hammer:2020dqp}: 
\begin{align}
 & S\left[g,A,\theta\right]=\int d^{4}x\sqrt{-g}\left[-\frac{R}{2\varkappa}+\right.\label{eq:axion_for_Lambda}\\
 & \left.+\frac{1}{2}\left(\partial\theta\right)^{2}+\frac{\theta}{f_{\Lambda}}F_{\alpha\beta}F^{\star\alpha\beta}-V_{\lambda}\left(\theta\right)\right]\,,\nonumber 
\end{align}
where now $\theta$ is a canonically normalized pseudoscalar and $f_{\Lambda}$
is some mass scale emulating axion decay constant and $V_{\lambda}\left(\theta\right)$
is a ``potential''. The presence of the standard kinetic term for
$\theta$ does not change anything, as on-shell $\theta$ is always
a constant $\theta_{\star}$ corresponding to the vacuum energy density
$V_{\lambda}$$\left(\theta_{\star}\right)$. In particular, this
action is parity symmetric. This gives a hope to find unimodular gravity
as a particular dynamical regime (maybe strongly coupled) of some
more usual high energy system. 

\section{Scale-Free Equations}

In this paper we propose a novel way to lose the trace part of the
Einstein equations. Indeed, instead of (\ref{eq:Unimod}) one can
write \emph{normalized} or \emph{scale-free }and still \emph{trace-trivial}
equations 
\begin{equation}
\frac{G_{\mu\nu}}{G}=\frac{T_{\mu\nu}}{T}\,,\label{eq:normalized}
\end{equation}
so that, the trace part of these normalized Einstein equations becomes
a useless identity $1=1$ instead of $0=0$ in the previous ``unimodular''
case. This \emph{unitrace} formulation does not require any dimensional
parameter, as both sides are dimensionless. If again $T_{\mu\nu}$
is covariantly conserved, one can follow the same procedure as for
the trace-free Einstein equations (\ref{eq:Unimod}). We apply covariant
derivative to both sides of (\ref{eq:normalized}) and use Bianchi
identity to obtain 
\begin{equation}
\frac{G_{\mu\nu}}{G^{2}}\nabla^{\mu}G=\frac{T_{\mu\nu}}{T^{2}}\nabla^{\mu}T\,.\label{eq:first_cons}
\end{equation}
Substituting here the normalized Einstein equations (\ref{eq:normalized})
one obtains instead of (\ref{eq:Consequence}) 
\begin{equation}
\partial_{\mu}\log\frac{G}{T}=0\,,\label{eq:Consequence_new}
\end{equation}
so that 
\begin{equation}
G=\beta T\,,\label{eq:relation_traces}
\end{equation}
where $\beta$ is constant of integration. Now we can substitute this
differential consequence (\ref{eq:relation_traces}) back into the
original normalized Einstein equations (\ref{eq:normalized}) to obtain
the standard Einstein equations 
\begin{equation}
G_{\mu\nu}=\beta T_{\mu\nu\,.}\label{eq:restored_Einstein}
\end{equation}

Thus, this novel trace-trivial formulation (\ref{eq:normalized})
allows to have the Newton constant 
\begin{equation}
G_{N}=\frac{\beta}{8\pi}\,,\label{eq:Newton_alpha}
\end{equation}
 as a constant of integration or global dynamical degree of freedom.
Similarly to $\Lambda$ in the ``unimodular'' case, this constant
of integration $\beta$ is allowed to be both positive and negative. 

Here a cautionary remark is necessary: not only we assumed that the
traces $T$ and $G$ are non-vanishing, but also we required a stronger
condition that $G_{\mu\nu}$ (or $T_{\mu\nu}$) has an inverse for
interesting solutions. Of course gravity waves in empty space would
violate this assumption. However, conformal anomaly can motivate the
non-vanishing trace of the EMT\footnote{In particular, the trace of total EMT and the corresponding Ricci
scalar are not entirely vanishing during the early universe radiation
domination époque, see e.g. \citep{Caldwell:2013mox}. }, while an existence of an arbitrary small cosmological constant (included
into this EMT) would imply desired invertibility for all solutions,
except of measure zero when this CC is exactly compensated by some
matter. 

Interestingly, one can obtain \emph{normalized}, \emph{scale-free
}and \emph{trace-trivial} equations (\ref{eq:normalized}) from a
restricted variation of the Einstein-Hilbert action. However, now
instead of the transversality condition (\ref{eq:transversality})
one should take either ``Einstein transversality'' condition: 
\begin{equation}
G_{\mu\nu}\bar{\delta}g^{\mu\nu}=0\,,\label{eq:Einstein_transvers}
\end{equation}
or the ``energy transversality'' condition: 
\begin{equation}
T_{\mu\nu}\bar{\delta}g^{\mu\nu}=0\,.\label{eq:Energy_transverse}
\end{equation}
These conditions are mutually complementary and interchangeable. Indeed,
assuming one of them results through the general unrestricted variational
relation (\ref{eq:unrestricted_variation}) in the other one. Either
of them would be deadly for unrestricted variations, but here we should
merely require equality of only those parts of $T_{\mu\nu}$ which
are not proportional to $G_{\mu\nu}$. Namely, imposing the ``Einstein
transversality'' condition (\ref{eq:unrestricted_variation}) yields
the Lagrange equations of the first kind 
\begin{equation}
\varkappa T_{\mu\nu}=\lambda\left(x\right)G_{\mu\nu}\,,\label{eq:first_kind_Einstein_transvers}
\end{equation}
where $\lambda\left(x\right)$ is a Lagrange multiplier. Now we can
again multiply both sides of this equation with $g^{\mu\nu}$ and
exclude the Lagrange multiplier to obtain (\ref{eq:normalized}).
Clearly, the ``energy transversality'' case is treated completely
analogously. Similarly to the unimodular gravity we could apply covariant
derivative and use Bianchi identity along with the \emph{assumed}
covariant conservation of EMT directly in (\ref{eq:first_kind_Einstein_transvers})
to find that $\lambda\left(x\right)=const$. 

\section{Many faces of Lagrange}

The proportionality between $G_{\mu\nu}$ and $T_{\mu\nu}$ in our
scale-free formulation, or proportionality between the metric $g_{\mu\nu}$
and $\left(G_{\mu\nu}-\varkappa T_{\mu\nu}\right)$ in the case of
the unimodular gravity, can be expressed without any use of the Lagrange
multiplier. One can follow an analogy with usual vectors. There, any
two vectors are proportional, provided their vector (wedge) product
vanishes and vice versa. Thus, for our novel scale-free gravity one
can write analogous expression
\begin{equation}
G_{\mu\nu}T_{\alpha\beta}=G_{\alpha\beta}T_{\mu\nu}\,,\label{eq:antisymmetric_G}
\end{equation}
instead of the Lagrange's equations of the first kind for the scale-free
formulation (\ref{eq:first_kind_Einstein_transvers}), while for the
unimodular case one would write 
\begin{equation}
\left(G_{\mu\nu}-\varkappa T_{\mu\nu}\right)g_{\alpha\beta}=\left(G_{\alpha\beta}-\varkappa T_{\alpha\beta}\right)g_{\mu\nu}\,,\label{eq:antisymmetric_CC}
\end{equation}
instead of (\ref{eq:1Kind_Lagrange_Unimod}). 

Further, it is noteworthy that one can exclude the Lagrange multipliers
from the Lagrange equations of the first kind (\ref{eq:1Kind_Lagrange_Unimod}),
(\ref{eq:first_kind_Einstein_transvers}) in many different ways.
The multiplication of (\ref{eq:1Kind_Lagrange_Unimod}), (\ref{eq:first_kind_Einstein_transvers})
with respect to the contravariant metric has a clear advantage that
this inverse metric always exists and that it's trace is always equal
to the number of spacetime dimensions. However, one could use other
equivalent ways to eliminate the Lagrange multiplier and write equations
in terms of known fields in different forms. 

\subsubsection{Unimodular gravity}

For example, in unimodular gravity one could multiply (\ref{eq:1Kind_Lagrange_Unimod})
with $G^{\mu\nu}$ and, under the assumption that $G\neq0$, obtain
\begin{equation}
\lambda=\frac{G^{\mu\nu}}{G}\left(G_{\mu\nu}-\varkappa T_{\mu\nu}\right)\,,\label{eq:unimod_G}
\end{equation}
so that 
\begin{equation}
G_{\mu\nu}-\varkappa T_{\mu\nu}=\frac{G^{\alpha\beta}}{G}\left(G_{\alpha\beta}-\varkappa T_{\alpha\beta}\right)g_{\mu\nu}\,.\label{eq:unimod_project_G}
\end{equation}
Or repeating the same procedure with $T_{\mu\nu}$ 
\begin{equation}
G_{\mu\nu}-\varkappa T_{\mu\nu}=\frac{T^{\alpha\beta}}{T}\left(G_{\alpha\beta}-\varkappa T_{\alpha\beta}\right)g_{\mu\nu}\,.\label{eq:unimod_project_T}
\end{equation}
Clearly, one could take any second rank tensor $M^{\alpha\beta}$
constructed from the matter fields and the metric and then use it
instead of $T^{\alpha\beta}$ or $G^{\alpha\beta}$, provided the
trace $M\neq0$. 

On the other hand, instead of taking the trace one can take the determinant
of (\ref{eq:1Kind_Lagrange_Unimod}) to obtain 
\begin{equation}
\lambda^{4}=\frac{\text{det}\left(G_{\mu\nu}-\varkappa T_{\mu\nu}\right)}{\text{det}g_{\alpha\beta}}\,.\label{eq:unimod_det_Lagrange}
\end{equation}
Consequently for positive $\lambda$ instead of trace-free Einstein
equations (\ref{eq:Unimod}) of the unimodular gravity one acquires
equations on unimodular parts of the corresponding tensors
\begin{equation}
\frac{G_{\mu\nu}-\varkappa T_{\mu\nu}}{\left(-\text{det}\left(G_{\alpha\beta}-\varkappa T_{\alpha\beta}\right)\right)^{1/4}}=\frac{g_{\mu\nu}}{\left(-\text{det}g_{\alpha\beta}\right)^{1/4}}\,.\label{eq:unimodular_relation}
\end{equation}
Of course, taking the trace of these equations one excludes determinants
and arrives back to trace-free Einstein equations (\ref{eq:Unimod}). 

\subsubsection{Scale-free gravity}

Similarly one can write the \emph{scale-free }Einstein equations\emph{
}(\ref{eq:normalized}) in many different ways. For example, analogously
to the unimodular case one can use determinant to solve for the Lagrange
multiplier in the unitrace case and write 
\begin{equation}
\frac{G_{\mu\nu}}{\left(\text{det}G_{\alpha\beta}\right)^{1/4}}=\frac{T_{\mu\nu}}{\left(\text{det}T_{\alpha\beta}\right)^{1/4}}\,,\label{eq:unimodular_for_G}
\end{equation}
so that again there is no scale - no dimensionful parameter. It looks
like it is a very different equation. However, taking the trace one
acquires that 
\begin{equation}
\frac{\left(\text{det}T_{\mu\nu}\right)^{1/4}}{\left(\text{det}G_{\mu\nu}\right)^{1/4}}=\frac{T}{G}\,,\label{eq:solving_traces}
\end{equation}
so that one can get rid of the ratio of the determinants and obtains
in this way the same (\ref{eq:normalized}). This different parametrization
can be as general as 
\begin{equation}
\frac{G_{\mu\nu}}{f\left(G_{\alpha\beta}\right)}=\frac{T_{\mu\nu}}{f\left(T_{\alpha\beta}\right)}\,,\label{eq:general_homgeneous_f}
\end{equation}
where $f$ is general homogeneous function of the first degree of
the tensor argument. In that case, there is no scale in this equation.
This function can also depend on the metric. Instead of the built-in
identity $1=1$ with respect to taking the trace\footnote{Note that the operation of taking trace can be considered as a particular
example of such function on tensors, $f\left(\mathcal{O}_{\alpha\beta}\right)=g^{\alpha\beta}\mathcal{O}_{\alpha\beta}$,
which is homogeneous of the first degree.} of (\ref{eq:normalized}), here the corresponding identity is $f\left(G_{\mu\nu}/f\left(G_{\alpha\beta}\right)\right)=1\,.$
However, taking the trace one gets back (\ref{eq:normalized}) again.
For example, 
\begin{equation}
\frac{G_{\mu\nu}}{\left(-g\right)^{1/4}G+\left(\text{det}G_{\alpha\beta}\right)^{1/4}}=\frac{T_{\mu\nu}}{\left(-g\right)^{1/4}T+\left(\text{det}T_{\alpha\beta}\right)^{1/4}}\,,\label{eq:strange_example}
\end{equation}
still reproduces exactly the same dynamics as (\ref{eq:normalized}).
Similarly to the unimodular case, one can also use $G^{\mu\nu}$ to
express the Lagrange multiplier in (\ref{eq:first_kind_Einstein_transvers})
as
\begin{equation}
\lambda\left(x\right)=\varkappa\frac{T_{\mu\nu}G^{\mu\nu}}{G^{\alpha\beta}G_{\alpha\beta}}\,.\label{eq:Lagrange_for_multiplying_with_G}
\end{equation}
In this way, instead of \emph{unitrace }equations\emph{ }(\ref{eq:normalized})
one arrives to scale-free equations 
\begin{equation}
T_{\mu\nu}=\frac{T_{\sigma\lambda}G^{\sigma\lambda}}{G^{\alpha\beta}G_{\alpha\beta}}\,G_{\mu\nu}\,,\label{eq:multiplying_with_G}
\end{equation}
with built-in identity with respect to multiplication with $G^{\mu\nu}$.
On the other hand, multiplying (\ref{eq:first_kind_Einstein_transvers})
with $T^{\mu\nu}$ one obtains 
\begin{equation}
\frac{T_{\sigma\lambda}G^{\sigma\lambda}}{T^{\alpha\beta}T_{\alpha\beta}}\,T_{\mu\nu}=G_{\mu\nu}\,.\label{eq:multiplying_with_T}
\end{equation}

Again, instead of $T_{\mu\nu}$ or $G_{\mu\nu}$ one could employ
any second rank tensor $M^{\alpha\beta}$ constructed from the matter
fields and metric, provided the trace $M^{\alpha\beta}G_{\alpha\beta}\neq0$,
so that 
\begin{equation}
\lambda\left(x\right)=\varkappa\frac{T_{\mu\nu}M^{\mu\nu}}{M^{\alpha\beta}G_{\alpha\beta}}\,,\label{eq:Lambda_M}
\end{equation}
which yields
\begin{equation}
T_{\mu\nu}=\frac{T_{\sigma\lambda}M^{\sigma\lambda}}{M^{\alpha\beta}G_{\alpha\beta}}\,G_{\mu\nu}\,.\label{eq:Einstein_Equation_M}
\end{equation}
In particular, one could even use inverse tensors $\left(T^{-1}\right)^{\mu\nu}$
or $\left(G^{-1}\right)^{\mu\nu}$ or $\left(R^{-1}\right)^{\mu\nu}$
: defined as $\left(T^{-1}\right)^{\mu\lambda}T_{\lambda\nu}=\delta_{\nu}^{\mu}$
etc., provided they exist\footnote{For a very recent discussion of other use of inverse Ricci tensor
or anti-curvature in cosmology, see \citep{Amendola:2020qho}.}. In this case, one would write 
\begin{equation}
G_{\mu\nu}=\frac{G_{\sigma\lambda}\left(T^{-1}\right)^{\lambda\sigma}}{4}\,T_{\mu\nu}\,,\label{eq:Einstein_Inverse_T}
\end{equation}
instead of (\ref{eq:normalized}). 

\section{Einstein-Transversality in Action for Scale-Free Gravity \label{sec:Action-for-Einstein-transverse} }

By analogy with the Henneaux-Teitelboim formulation of the unimodular
gravity (\ref{eq:vector_for_CC}), it is easy to write similar actions
for a global degree of freedom representing the Newton constant: 
\begin{equation}
S\left[g,C,\alpha\right]=\frac{1}{2}\int d^{4}x\sqrt{-g}\left(\nabla_{\mu}C^{\mu}-R\right)\alpha\,.\label{eq:Vector_for_G}
\end{equation}
An action similar to (\ref{eq:Vector_for_G}) was used in \citep{Kaloper:2015jra,Kaloper:2018kma}
for local version of the vacuum energy sequester \citep{Kaloper:2013zca}.
On the other hand, following (\ref{eq:Chern_Simons_CC}) one can exchange
the vector $C^{\mu}$ with the Chern-Simons current of a gauge field
$\mathcal{A}_{\mu}$ and write 
\begin{equation}
S\left[g,\mathcal{A},\alpha\right]=\frac{1}{2}\int d^{4}x\sqrt{-g}\left(\mathcal{F}_{\mu\nu}\mathcal{F}^{\star\mu\nu}-R\right)\alpha\,.\label{eq:Chern_Simons_for_G}
\end{equation}
Here we assume standard general-covariant and minimally coupled action
for all usual matter fields. 

Checking the equations of motion for (\ref{eq:Vector_for_G}) we have
\begin{equation}
\frac{2}{\sqrt{-g}}\cdot\frac{\delta S}{\delta C^{\mu}}=-\partial_{\mu}\alpha=0\,,\label{eq:C_eom}
\end{equation}
and for the metric variation 
\begin{align}
 & \delta_{g}S=\frac{1}{2}\int d^{4}x\left[-\alpha\delta_{g}\left(\sqrt{-g}R\right)+\alpha\delta_{g}\partial_{\mu}\left(\sqrt{-g}C^{\mu}\right)\right]=\nonumber \\
 & =\frac{1}{2}\int d^{4}x\left[-\alpha\delta_{g}\left(\sqrt{-g}R\right)-\left(\delta\sqrt{-g}\right)C^{\mu}\partial_{\mu}\alpha\right]\,,\label{eq:variation_C_for_G}
\end{align}
so that on equation of motion for $C^{\mu}$ the divergence term does
not contribute to the metric equation of motion. In this way, for
the standard action of the minimally coupled matter fields one obtains
\begin{equation}
\alpha\,G_{\mu\nu}=T_{\mu\nu}\,,\label{eq:alpha_Einstein}
\end{equation}
and consequently 
\begin{equation}
\alpha=\frac{1}{8\pi G_{N}}\,.\label{eq:G_N_alpha}
\end{equation}
On the other hand, $\sqrt{-g}\mathcal{F}_{\mu\nu}\mathcal{F}^{\star\mu\nu}$
does not contribute to the tensor equations of motion, see (\ref{eq:dual}).
Whereas for abelian $\mathcal{A}_{\beta}$ 
\begin{equation}
\frac{2}{\sqrt{-g}}\cdot\frac{\delta S}{\delta\mathcal{A}_{\gamma}}=4\mathcal{F}^{\star\gamma\mu}\partial_{\mu}\alpha=0\,.\label{eq:equation_gauge_field-1}
\end{equation}
However, to conclude here that $\partial_{\mu}\alpha=0$ one has to
assume the existence of $\left(\mathcal{F}^{\star\gamma\mu}\right)^{-1}$.
The Hodge dual of the field tensor has an inverse provided $\mathcal{F}_{\mu\nu}\mathcal{F}^{\star\mu\nu}\neq0$,
for a recent discussion see \citep{DeFelice:2019hxb}, hence one should
assume that $R$ is not identically vanishing\footnote{It is worth noting here again that in the early universe $R$ is not
vanishing \citep{Caldwell:2013mox}.}. 

The canonical structure is more transparent for the action (\ref{eq:Vector_for_G}).
There one can either follow a direct analogy with the Henneaux and
Teitelboim formulation of the unimodular gravity \citep{Henneaux:1989zc},
and Faddeev-Jackiw \citep{Faddeev:1988qp,Jackiw:1993in} procedure
or look at the Hamiltonian analysis performed for the vacuum energy
sequester, see \citep{Bufalo:2016omb,Kluson:2014tma,Svesko:2018cbo}.
Similarly to (\ref{eq:cosmic_time}), in this action $\alpha$ is
a canonical momentum conjugated to the global quantity 
\begin{equation}
\mathscr{R}\left(t\right)=\frac{1}{2}\int_{\Sigma}d^{3}\mathbf{x}\sqrt{-g}\,C^{t}\left(t,\mathbf{x}\right)\,,\label{eq:global_curvature_charge}
\end{equation}
which measures integrated Ricci scalar between the Cauchy hypersurfaces
$\Sigma_{2}$ and $\Sigma_{1}$
\begin{equation}
\mathscr{R}\left(t_{2}\right)-\mathscr{R}\left(t_{1}\right)=\frac{1}{2}\int_{\mathscr{V}}d^{4}x\sqrt{-g}\,R\,.\label{eq:4volume-1}
\end{equation}
Again one can think about $\mathscr{R}$ as a charge which is not
conserved, but rather continuously produced by Ricci curvature which
plays the role of a source 
\begin{equation}
\nabla_{\mu}C^{\mu}=R\,.\label{eq:source_for_G}
\end{equation}
Thus, the inverse Newton constant $\alpha$ is canonically conjugated\footnote{For thermodynamical arguments in favor of this relation, see \citep{Volovik:2020qtp,Klinkhamer:2008ff}. }
to the spacetime averaged Ricci curvature $\mathscr{R}\left(t\right)$.
On the other hand, $\mathscr{R}$ is clearly proportional to the Einstein-Hilbert
(volume) part of the gravitational action. Thus, this resembles action-angle
variables. As $\alpha$ is canonically conjugated to $\mathscr{R}$
one can directly apply the Heisenberg uncertainty relation to obtain
\begin{equation}
\frac{\delta G_{N}}{G_{N}}\times\frac{\delta\int_{\mathscr{V}}d^{4}x\sqrt{-g}\,R}{\ell_{Pl}^{2}}\geq8\pi\,.\label{eq:uncertainty_G_alpha}
\end{equation}
It is instructive to express this inequality in terms of fluctuations
of the corresponding Planck length $\ell_{Pl}=\sqrt{\hbar G_{N}}$
\begin{equation}
\frac{\delta\ell_{Pl}}{\ell_{Pl}}\times\frac{\delta\int_{\mathscr{V}}d^{4}x\sqrt{-g}\,R}{\ell_{Pl}^{2}}\geq4\pi\,.\label{eq:uncertainty_Planck_length}
\end{equation}
Similarly to (\ref{eq:uncertainty_volume}), these inequalities may
have nontrivial consequences close to singularities. Interestingly,
$\hbar$ is not explicit in these uncertainty relations above. 

Following discussion around equations (\ref{eq:W_transform}), (\ref{eq:fixed_W})
and (\ref{eq:fixed_W_constraint}) and applying them to $C^{\mu}$
instead of $W^{\mu}$ one can try to fix
\begin{equation}
C^{\mu}=\delta_{t}^{\mu}\frac{t}{\sqrt{-g}}\,M_{\alpha}^{2}\,,\label{eq:fixed_C}
\end{equation}
in the action, \emph{before} variation. Here $M_{\alpha}$ is some
mass scale introduced for dimensional reasons. Then, in units where
$M_{\alpha}=1$, the action takes a non-covariant form
\begin{equation}
S\left[g,\alpha\right]=\frac{1}{2}\int d^{4}x\left(1-\sqrt{-g}R\right)\alpha\,,\label{eq:unicurvature_action}
\end{equation}
thus instead of the ``unimodular'' constraint (\ref{eq:fixed_W_constraint})
we have
\begin{equation}
\sqrt{-g}R=1\,,\label{eq:unicurvature}
\end{equation}
or the ``unicurvature'' condition. Similarly to what is often done
in the unimodular gravity one can forget about gauge and coordinate
fixing steps resulting in (\ref{eq:fixed_C}) and take this non-covariant
action (\ref{eq:unicurvature_action}), as a starting point. 

Following a nice discussion in \citep{vanderBij:1981ym}, one can
always construct such coordinates where (\ref{eq:unicurvature}) is
fulfilled. Suppose in some coordinates $\left(t',x'^{i}\right)$ we
have $\sqrt{-g'}R'\neq1$. Then using that the Ricci density transforms
as
\begin{equation}
\sqrt{-g'}R'=\left|\frac{\partial x}{\partial x'}\right|\sqrt{-g}R\,,\label{eq:Ricci_density_transform}
\end{equation}
we can find such coordinates $\left(t,x^{i}\right)$ that $x'^{i}=x^{i}$,
but 
\begin{equation}
t=\int^{t'}dt''\sqrt{-g'\left(t'',x'^{i}\right)}\,R'\left(t'',x'^{i}\right)\,,\label{eq:new_time_unicurv}
\end{equation}
so that one obtains
\begin{equation}
\left|\frac{\partial x}{\partial x'}\right|=\frac{\partial t}{\partial t'}=\sqrt{-g'}R'\,,\label{eq:det_transform_t}
\end{equation}
which enforces the ``unicurvature'' condition (\ref{eq:unicurvature})
in coordinates $\left(t,x^{i}\right)$. However, similarly to the
non-covariant formulation of unimodular gravity one cannot just find
coordinates where both conditions (\ref{eq:fixed_C}) and (\ref{eq:unicurvature})
hold simultaneously. Therefore the latter condition is an equation
of motion implying nontrivial dynamics. 

Let us check that, despite the illegitimate step of fixing the gauge
directly in the action before variation, still the action (\ref{eq:unicurvature_action})
reproduces the same dynamics as the original general-covariant action
(\ref{eq:Vector_for_G}). Variation with respect to $\delta g^{\mu\nu}$
of (\ref{eq:unicurvature_action}) accompanied with the action for
minimally coupled matter yields 
\begin{equation}
\alpha G_{\mu\nu}-\nabla_{\mu}\nabla_{\nu}\alpha+g_{\mu\nu}\Box\alpha=T_{\mu\nu}\,,\label{eq:unicurvature_eom_alpha}
\end{equation}
where $\Box=g^{\mu\nu}\nabla_{\mu}\nabla_{\nu}$. Here we are obviously
missing (\ref{eq:C_eom}) which enforces that $\alpha$ is a constant.
However, applying $\nabla^{\mu}$ to both sides of the equation (\ref{eq:unicurvature_eom_alpha})
one obtains 
\begin{equation}
G_{\mu\nu}\nabla^{\mu}\alpha+\nabla_{\nu}\Box\alpha-\Box\nabla_{\nu}\alpha=0\,,\label{eq:Bianchi_result_alpha}
\end{equation}
where we used the Bianchi identity and EMT conservation. Further,
invoking the identity 
\begin{equation}
\nabla_{\nu}\Box\alpha-\Box\nabla_{\nu}\alpha=-R_{\mu\nu}\nabla^{\mu}\alpha\,,\label{eq:commutator_identity}
\end{equation}
yields $R\partial_{\nu}\alpha=0,$ so that we are back to (\ref{eq:alpha_Einstein}).

Interestingly, varying the ``unicurvature'' constraint (\ref{eq:unicurvature})
one obtains 
\begin{equation}
G_{\mu\nu}\bar{\delta}g^{\mu\nu}=\nabla_{\mu}\nabla_{\nu}\bar{\delta}g^{\mu\nu}-g_{\mu\nu}\Box\bar{\delta}g^{\mu\nu}\,,\label{eq:Modified_Einstein_trans}
\end{equation}
reproducing the ``Einstein transversality'' condition (\ref{eq:Einstein_transvers})
only up to a divergence. 

It is worth mentioning, that similarly to what is sometimes done in
the unimodular gravity one can also substitute $\sqrt{-g}=R^{-1}$
into matter part of the action. This would produce an apparently new
and crazily looking theory without changing any physics. 

Finally, we would like to mention that it is the absence of the usual
kinetic term $\mathcal{F}_{\mu\nu}\mathcal{F}^{\mu\nu}$ which forces
$\alpha$ to be constant. Hence, similarly to (\ref{eq:axion_for_Lambda}),
one can write 
\begin{align}
 & S\left[g,\mathcal{A},\nu\right]=\int d^{4}x\sqrt{-g}\left[-\frac{1}{2}\nu^{2}R+\right.\label{eq:axion_for_Newton}\\
 & \left.+\frac{1}{2}\left(\partial\nu\right)^{2}+\frac{\nu}{f_{\alpha}}\mathcal{F}_{\gamma\sigma}\mathcal{F}^{\star\gamma\sigma}-V_{\alpha}\left(\nu\right)\right]\,,\nonumber 
\end{align}
where now $\nu$ is canonically normalized pseudoscalar and $f_{\alpha}$
is some mass scale emulating the axion would-be decay constant, while
$V_{\alpha}\left(\nu\right)$ is a potential. Again, the presence
of the standard kinetic term for $\nu$ does not change anything,
as on-shell $\nu$ always remains an arbitrary constant $\nu_{\star}$.
In particular, the kinetic term may be non-canonical. Another advantage
is that this action has usual properties regarding parity transformations.
The action is written in the Jordan frame. The potential induces cosmological
constant $V_{\alpha}\left(\nu_{\star}\right)$ whose presence is also
useful for invertibility of $\mathcal{F}^{\star\alpha\beta}$. This
construction gives hope to find the \emph{scale-free} gravity (\ref{eq:normalized})
as a particular dynamical regime (maybe strongly coupled) of a more
usual QFT system. 

\section{Energy-Transversality in Action for Scale-Free Gravity\label{sec:Action-for-energy-transverse}}

On the other hand, one can attain the effective Newton constant as
a global degree of freedom by changing the usual action for matter
fields $\Phi_{m}$, 
\begin{equation}
S_{0}\left[g,\Phi_{m}\right]=\int d^{4}x\sqrt{-g}\,\mathscr{L}_{m}\,,\label{eq:usual_Action_matter}
\end{equation}
to the one (c.f. \citep{Carroll:2017gqo}) similar to the Henneaux
and Teitelboim construction (\ref{eq:vector_for_CC}) 
\begin{equation}
S\left[g,\beta,L,\Phi_{m}\right]=\int d^{4}x\sqrt{-g}\,\beta\left(\mathscr{L}_{m}-\nabla_{\lambda}L^{\lambda}\right)\,.\label{eq:rescaled_matter_action}
\end{equation}
Or instead of the vector field $L^{\lambda}$ one can again use a
gauge field $\mathscr{A}_{\mu}$
\begin{equation}
S\left[g,\beta,\mathscr{A},\Phi_{m}\right]=\int d^{4}x\sqrt{-g}\,\beta\left(\mathscr{L}_{m}-\mathscr{F}_{\mu\nu}\mathscr{F}^{\star\mu\nu}\right)\,,\label{eq:rescaled_matter_axion}
\end{equation}
so that the ``axionic'' field $\beta$ \emph{universally} couples
to \emph{all} matter fields. Here we assume that the gravitational
sector is described by the standard Einstein-Hilbert action. We could
slightly modify the action in order to have better transformation
properties with respect to parity symmetry and write 
\begin{equation}
S\left[g,\eta,\mathscr{A},\Phi_{m}\right]=\int d^{4}x\sqrt{-g}\,\eta\left(\eta\mathscr{L}_{m}-\mathscr{F}_{\mu\nu}\mathscr{F}^{\star\mu\nu}\right)\,.\label{eq:P_improved_action}
\end{equation}
Under parity $\eta$ (and $\beta$) is a pseudoscalar which transforms
as $\eta\rightarrow-\eta$. Lest the normal matter become ghosty under
parity transformation, one should use this improved version of the
action (\ref{eq:P_improved_action}). 

Of course, completely analogously to previous cases, the variation
with respect to the vector field $L^{\mu}$ (or $\mathscr{A}_{\mu}$)
results in $\partial_{\mu}\beta=0$ (or $\partial_{\mu}\eta=0$).
On shell $\eta=\eta_{\star}$, which opens a path to further modification
of this action to make it look more familiar 
\begin{align}
 & S\left[g,\eta,\mathscr{A},\Phi_{m}\right]=\int d^{4}x\sqrt{-g}\left[\frac{1}{2}\left(\partial\eta\right)^{2}-V_{\beta}\left(\eta\right)+\right.\nonumber \\
 & \qquad\qquad\qquad\left.+\frac{\eta^{2}}{M_{m}^{2}}\mathscr{L}_{m}-\frac{\eta}{f_{\beta}}\mathscr{F}_{\mu\nu}\mathscr{F}^{\star\mu\nu}\right]\,,\label{eq:improvement_by_kinetic}
\end{align}
where $V_{\beta}\left(\eta\right)$ is a potential and the usual dimensions
of $\eta$ are restored by introduction of two mass scales $M_{m}$
and $f_{\beta}$. And again the presence of the potential introduces
a cosmological constant $V_{\beta}\left(\eta_{\star}\right)$. This
guarantees the existence of $\left(\mathscr{F}^{\star\mu\nu}\right)^{-1}$
required to infer the constancy of $\eta$. It is crucial that the
axion \emph{$\eta$ }is\emph{ universally} coupled to all other matter
fields. While gravity is assumed to be described by the standard Einstein-Hilbert
action so that the description is in the Einstein frame. Clearly,
in our setup the mass scales and even the form of the potential $V_{\beta}\left(\eta\right)$
and the form of the kinetic term (which can be again non-canonical)
are rather irrelevant. However, in any attempt to embed this theory
into more usual QFT setup these will be crucial bits of information.
This action appears from standard Yang-Mills construction provided
one takes the limit when one can neglect the usual kinetic term $-\mathscr{F}_{\mu\nu}\mathscr{F}^{\mu\nu}/4\text{g}^{2}$
for the gauge field. Naively, this limit corresponds to a confinement
or an infinitely strong coupling $\text{g}\rightarrow\infty$. 

It is worth noting that combining action (\ref{eq:improvement_by_kinetic})
or (\ref{eq:axion_for_Newton}) with the (\ref{eq:axion_for_Lambda})
provides an axionic formulation of the vacuum energy sequestering
\citep{Kaloper:2013zca,Kaloper:2015jra,Padilla:2015aaa}. This novel
axionic formulation of sequester is much closer to the usual particle
physics models. 

The canonical structure of the system is more transparent for the
action (\ref{eq:rescaled_matter_action}). In this formulation of
the theory, on the right hand side of the Einstein equations one obtains
a rescaled stress tensor for matter fields 
\begin{equation}
T_{\mu\nu}=\frac{2}{\sqrt{-g}}\frac{\delta S}{\delta g^{\mu\nu}}=\beta\,T_{\mu\nu}^{\left(m\right)},\label{eq:rescaled_EMT}
\end{equation}
where 
\begin{equation}
T_{\mu\nu}^{\left(m\right)}=\frac{2}{\sqrt{-g}}\frac{\delta S_{0}}{\delta g^{\mu\nu}}=\frac{2}{\sqrt{-g}}\frac{\delta\left(\sqrt{-g}\mathscr{L}_{m}\right)}{\delta g^{\mu\nu}}\,,\label{eq:usual_EMT_matter}
\end{equation}
is defined exclusively through the usual Lagrangian density or from
the original action (\ref{eq:usual_Action_matter}). In the last equality
we assumed that matter sector is standard and in particular that the
Lagrangian does not depend on the derivatives of the metric. In fact,
this last equality would not be applicable for \emph{Kinetic Gravity
Braiding} \citep{Deffayet:2010qz} and \emph{G-Inflation} \citep{Kobayashi:2010cm}. 

Similarly to (\ref{eq:cosmic_time}) and (\ref{eq:global_curvature_charge})
the dynamical degree of freedom canonically conjugated to $\beta$
is 
\begin{equation}
I\left(t\right)=-\int_{\Sigma}d^{3}\mathbf{x}\sqrt{-g}\,L^{t}\left(t,\mathbf{x}\right)\,,\label{eq:Matter_action_charge}
\end{equation}
which measures the usual matter action between the Cauchy hypersurfaces
$\Sigma_{2}$ and $\Sigma_{1}$
\begin{equation}
I\left(t_{2}\right)-I\left(t_{1}\right)=-\int_{\mathscr{V}}d^{4}x\sqrt{-g}\,\mathscr{L}_{m}\,.\label{eq:Matter_Action}
\end{equation}
Also here one can think about $I\left(t\right)$ as a charge which
is not conserved, but rather continuously produced by the total matter
Lagrangian which plays the role of a source 
\begin{equation}
\nabla_{\mu}L^{\mu}=\mathscr{L}_{m}\,.\label{eq:creating_matter_action}
\end{equation}
On the other hand, the introduction of $\beta$ rescales the commutation
relations for usual matter. In the same way, in the approach with
the Einstein-transverse condition described in (\ref{sec:Action-for-Einstein-transverse})
a similar rescaling (\ref{eq:rescaled_canonical_commutator}) with
$1/G_{N}$ instead of $\beta$ occurs for the commutator of the \emph{dimensionless}
spatial metric and it's conjugated momentum given by the trace-free
extrinsic curvature density, see \citep{Poisson}. 

For example, for the usual scalar field $\phi$ the canonical momentum
gets rescaled in the analogous way to the EMT: 
\begin{equation}
\pi=\beta\pi^{\left(m\right)}=\beta\sqrt{-g}\frac{\partial\mathscr{L}_{m}}{\partial\dot{\phi}}\,.\label{eq:canonical_momentum_rescaled}
\end{equation}
Hence, if the canonical commutator is 
\begin{equation}
\left[\phi\left(\mathbf{x}\right),\pi\left(\mathbf{y}\right)\right]=i\hbar\delta\left(\mathbf{x}-\mathbf{y}\right)\,,\label{eq:canonical_commutator}
\end{equation}
the usual commutator gets rescaled as 
\begin{equation}
\left[\phi\left(\mathbf{x}\right),\pi^{\left(m\right)}\left(\mathbf{y}\right)\right]=\frac{i\hbar}{\beta}\delta\left(\mathbf{x}-\mathbf{y}\right)\,,\label{eq:rescaled_canonical_commutator}
\end{equation}
so that the effective Planck constant is 
\begin{equation}
\bar{\hbar}=\frac{\hbar}{\beta}\,.\label{eq:rescaled_Planck_constant}
\end{equation}
This is ideologically similar to what is discussed in \citep{Carroll:2017gqo},
though our frameworks are different. 

For the usual Einstein-Hilbert action with fixed $G_{N}$ one obtains
rescaled Einstein equations 
\begin{equation}
\frac{G_{\mu\nu}}{8\pi G_{N}}=\beta T_{\mu\nu}^{\left(m\right)}\,,\label{eq:beta_Einstein}
\end{equation}
so that the effective Newton constant, $\bar{G}_{N}$, is rescaled
as
\begin{equation}
\bar{G}_{N}=G_{N}\beta\,.\label{eq:rescaled_G_N}
\end{equation}
Interestingly, one obtains 
\begin{equation}
\bar{\hbar}\bar{G}_{N}=\hbar G_{N}\,,\label{eq:invariance}
\end{equation}
so that the Planck length (and time) $\ell_{Pl}=\sqrt{\hbar G_{N}}$
remain invariant. This discussion is applicable if the universe is
in an eigenstate of $\beta$. 

On the other hand, in generic state, $\beta$ can have its own quantum
fluctuations. In that case, we again write the Heisenberg uncertainty
relation for the fluctuation of the effective Newton constant (\ref{eq:rescaled_G_N})
\begin{equation}
\delta\bar{G}_{N}\times\delta\int_{\mathscr{V}}d^{4}x\sqrt{-g}\,\mathscr{L}_{m}\geq\frac{1}{2}\ell_{Pl}^{2}\,,\label{eq:matter_uncertainty_Realtion}
\end{equation}
or equivalently for the effective Planck constant (\ref{eq:rescaled_Planck_constant})
\begin{equation}
\delta\bar{\hbar}\times\delta\int_{\mathscr{V}}d^{4}x\sqrt{-g}\,\mathscr{L}_{m}\geq\frac{1}{2}\bar{\hbar}^{2}\,.\label{eq:uncertainty_for_Planck}
\end{equation}

We should remark that in case of the action (\ref{eq:P_improved_action})
instead of $\beta$ one should use $\eta^{2}$ in all rescaling above,
while for (\ref{eq:improvement_by_kinetic}) one uses $\eta^{2}/M_{m}^{2}$. 

Interestingly, the invariance of $\ell_{Pl}$ given by (\ref{eq:invariance})
does not allow for fluctuations of $\ell_{Pl}$, contrary to those
(\ref{eq:uncertainty_Planck_length}) present in the approach with
the Einstein-transverse condition. 

One can also combine both actions (\ref{eq:rescaled_matter_action})
and (\ref{eq:Vector_for_G}) and obtain the Einstein equations written
as 
\begin{equation}
\alpha\,G_{\mu\nu}=\beta\,T_{\mu\nu}^{\left(m\right)}\,,\label{eq:ab_Einstein}
\end{equation}
so that the effective Newton constant is 
\begin{equation}
\bar{G}_{N}=\frac{1}{8\pi}\,\frac{\beta}{\alpha}\,.\label{eq:G_effb/a}
\end{equation}

\subsection{\textquotedblleft Noncovariant\textquotedblright{} or \textquotedblleft unimatter\textquotedblright{}
formulation. }

Finally, similarly to fixing $W^{\mu}$ as in Eq.~(\ref{eq:fixed_W_constraint})
for the non-covariant formulation of the unimodular gravity and to
fixing $C^{\mu}$ as in Eq.~(\ref{eq:fixed_C}) for the non-covariant
``unicurvature'' gravity, one can try to illegitimately fix 
\begin{equation}
L^{\mu}=\delta_{t}^{\mu}\frac{t}{\sqrt{-g}}\,M_{\beta}^{4}\,,\label{eq:fixed_L}
\end{equation}
\emph{before} varying the action (\ref{eq:rescaled_matter_action}).
Here $M_{\beta}$ is some mass scale. Then, again in units where this
mass scale is unity, $M_{\beta}=1$, the action takes an unusual non-covariant
form 
\begin{equation}
S\left[g,\beta,\Phi_{m}\right]=\int d^{4}x\,\beta\left(\sqrt{-g}\mathscr{L}_{m}-1\right)\,,\label{eq:unimatter}
\end{equation}
implying unity of the matter Lagrangian density, which is clearly
different from Eq.$\,$(4) from \citep{Carroll:2017gqo}. On top,
we again assume the usual Einstein-Hilbert action with fixed $G_{N}$.
One could call this ``unimatter gravity'', as 
\begin{equation}
\sqrt{-g}\mathscr{L}_{m}=1\,,\label{eq:unimatter_constrain}
\end{equation}
which corresponds to a coordinate choice or a partial gauge fixing,
modulo the point which we have already discussed after Eq.~(\ref{eq:fixed_W_constraint}).
We can again repeat the same procedure to find proper coordinates
where ``unimatter'' condition holds by applying formulas (\ref{eq:Ricci_density_transform}),
(\ref{eq:new_time_unicurv}) and (\ref{eq:det_transform_t}) with
exchanging $R$ with $\mathscr{L}_{m}$. 

From the usual definition of the EMT (\ref{eq:usual_EMT_matter})
we have 
\begin{equation}
\delta\left(\sqrt{-g}\mathscr{L}_{m}\right)=\frac{\sqrt{-g}}{2}\,T_{\mu\nu}^{\left(m\right)}\bar{\delta}g^{\mu\nu}=0\,,\label{eq:Connecting_unmatter_energy_trans}
\end{equation}
thus the ``unimatter'' condition (\ref{eq:unimatter_constrain})
reproduces the ``energy transversality'' condition (\ref{eq:Energy_transverse})
restricting the variations $\bar{\delta}g^{\mu\nu}$. 

If, as it is the case for the Standard Model fields, $\mathscr{L}_{m}$
does not contain derivatives of the metric, then the variation with
respect to the metric yields (\ref{eq:beta_Einstein}). However, contrary
to the general-covariant formulation, now $\beta\left(x\right)$ is
a free function similarly to (\ref{eq:first_kind_Einstein_transvers}),
as we do not have an equation of motion for $L^{\mu}$ to impose $\partial_{\mu}\beta=0$.
In our restricted variational formulation with energy-transversality
condition (\ref{eq:Energy_transverse}) it was the joint work of the
Bianchi identity and the \emph{assumed} conservation of the EMT which
resulted in $\partial_{\mu}\beta=0$. However, similarly to what happens
\citep{Buchmuller:1988wx} (c.f. \citep{Percacci:2017fsy}) in the
non-covariant formulation (\ref{eq:fixed_W_constraint}) of the unimodular
gravity, due to the non-covariant formulation one cannot just assume
the conservation of $T_{\mu\nu}^{\left(m\right)}$. 

To illustrate this point, lets first consider just one scalar field
$\phi$, as general as k-essence \citep{ArmendarizPicon:2000ah,Chiba:1999ka}.
This can be also considered as a proxy for irrotational hydrodynamics.
The Euler-Lagrange equation for $\phi$ from the action (\ref{eq:unimatter})
is 
\begin{equation}
\sqrt{-g}\beta\frac{\partial\mathscr{L}_{m}}{\partial\phi}=\partial_{\mu}\left(\sqrt{-g}\beta\frac{\partial\mathscr{L}_{m}}{\partial\partial_{\mu}\phi}\right)\,,\label{eq:Euler_Lagrange_phi}
\end{equation}
and can be rewritten as 
\begin{equation}
\nabla_{\mu}\left(\frac{\partial\mathscr{L}_{m}}{\partial\partial_{\mu}\phi}\right)-\frac{\partial\mathscr{L}_{m}}{\partial\phi}=-\frac{\partial\mathscr{L}_{m}}{\partial\partial_{\mu}\phi}\frac{\partial_{\mu}\beta}{\beta}\,.\label{eq:beta_source}
\end{equation}
Further, for the scalar field the Noether form of the EMT is already
symmetric and can be written as 
\begin{equation}
T_{\quad\beta}^{\alpha\left(m\right)}=\frac{\partial\mathscr{L}_{m}}{\partial\partial_{\alpha}\phi}\partial_{\beta}\phi-\mathscr{L}_{m}\delta_{\beta}^{\alpha}\,,\label{eq:EMT_phi}
\end{equation}
while its divergence is 
\begin{equation}
\nabla_{\alpha}T_{\quad\beta}^{\alpha\left(m\right)}=\left(\nabla_{\alpha}\left(\frac{\partial\mathscr{L}_{m}}{\partial\partial_{\alpha}\phi}\right)-\frac{\partial\mathscr{L}_{m}}{\partial\phi}\right)\partial_{\beta}\phi\,.\label{eq:EMT_phi_div}
\end{equation}
In the usual case, the expression in front of $\partial_{\beta}\phi$
would be the equation of motion so that the EMT would be conserved.
However, here the equation of motion (\ref{eq:beta_source}) is modified
by the presence of $\beta$ so that there is a non-conservation:
\begin{equation}
\nabla_{\alpha}T_{\quad\mu}^{\alpha\left(m\right)}=-\frac{\partial\mathscr{L}_{m}}{\partial\partial_{\alpha}\phi}\,\frac{\partial_{\alpha}\beta}{\beta}\,\partial_{\mu}\phi\,.\label{eq:nonconservation}
\end{equation}
Now we can apply covariant derivative to the Einstein equations (\ref{eq:beta_Einstein})
and obtain 
\begin{align}
 & \frac{\nabla_{\alpha}G_{\mu}^{\alpha}}{8\pi G_{\text{N}}}=\nabla_{\alpha}\left(\beta T_{\quad\mu}^{\alpha\left(m\right)}\right)=\nonumber \\
 & =T_{\quad\mu}^{\alpha\left(m\right)}\partial_{\alpha}\beta+\beta\nabla_{\alpha}T_{\quad\mu}^{\alpha\left(m\right)}=-\mathscr{L}_{m}\partial_{\mu}\beta\,,\label{eq:div_Einstein_beta}
\end{align}
where we used (\ref{eq:nonconservation}) and (\ref{eq:EMT_phi}).
Hence, the Bianchi identity, indeed, again implies that $\beta$ is
a constant. This, of course, also results in the conservation of $T_{\mu\nu}^{\left(m\right)}$. 

Now let us consider usual free electrodynamics\footnote{supplemented by a cosmological constant $\Lambda$ to avoid that $\mathscr{L}_{m}=0$
on electromagnetic waves which are solutions of the free electrodynamics.} 
\begin{equation}
\mathscr{L}_{m}=-\frac{1}{4}F_{\mu\nu}F^{\mu\nu}+\Lambda\,,\label{eq:EM_Lagrangian}
\end{equation}
so that 
\begin{equation}
\frac{\partial\mathscr{L}_{m}}{\partial\partial_{\mu}A_{\nu}}=-F^{\mu\nu}\,,\label{eq:momentum_E}
\end{equation}
and we have the same equations of motion (\ref{eq:beta_source}),
where just instead of $\phi$ we have $A_{\mu}$, so that 
\begin{equation}
\nabla_{\lambda}F^{\mu\lambda}=-F^{\mu\lambda}\frac{\partial_{\lambda}\beta}{\beta}\,.\label{eq:divF}
\end{equation}
The symmetric Belinfante-Rosenfeld / Hilbert EMT relevant for gravity
\begin{equation}
T_{\quad\nu}^{\mu\left(m\right)}=F_{\:\,\lambda}^{\mu}F_{\:\,\nu}^{\lambda}-\mathscr{L}_{m}\delta_{\nu}^{\mu}\,,\label{eq:EM_emt}
\end{equation}
deviates from the canonical Noether one (\ref{eq:EMT_phi}) which
is not even covariantly conserved in the presence of gravity. In the
absence of the electromagnetic current one obtains 
\begin{equation}
\nabla^{\mu}T_{\mu\nu}^{\left(m\right)}=-F_{\lambda\nu}\nabla_{\mu}F^{\lambda\mu}=F_{\lambda\nu}F^{\lambda\alpha}\frac{\partial_{\alpha}\beta}{\beta}\,.\label{eq:divT_em_beta}
\end{equation}
Substituting this expression for non-conservation of EMT (\ref{eq:divT_em_beta})
along with (\ref{eq:EM_emt}) into the left hand side of (\ref{eq:div_Einstein_beta})
we again obtain the last equality there which, in turn, implies $\beta=const$. 

In both examples above we essentially used explicit forms (\ref{eq:EMT_phi})
and (\ref{eq:EM_emt}) of the EMT. The case with the scalar field
is easily generalizable to many scalar fields, but this is not a generic
matter source. Unfortunately, for generic matter with internal spin,
the canonical Noether EMT is not symmetric and is not a source of
gravity and does not coincide with the proper Hilbert EMT obtained
through the variation of the action with respect to the metric. Below
we provide the proof of the last equality in (\ref{eq:div_Einstein_beta})
for generic usual matter, whose Lagrangian does not involve derivatives
of the metric. However, this proof does not cover fermions, as we
do not want to overcomplicate the discussion by involving tetrads. 

The ``unimatter'' action (\ref{eq:unimatter}) is written in a particular
class of coordinates where (\ref{eq:fixed_L}) holds. Let us denote
one particular system of such coordinates $X^{\mu}$. Now we can write
this action (\ref{eq:unimatter}) in arbitrary coordinates $y^{\mu}$
by noting that $\mathscr{L}_{m}$, $d^{4}x\sqrt{-g}$ and $\beta$
are scalar objects:
\begin{equation}
S_{X}\left[g,\beta,\Phi_{m}\right]=\int d^{4}y\,\beta\left(\sqrt{-g}\mathscr{L}_{m}-\left|\frac{\partial X}{\partial y}\right|\right)\,,\label{eq:unimatter_Jacobian}
\end{equation}
where $\left|\partial X/\partial y\right|$ is the Jacobian. We should
stress that in this action the original coordinates $X^{\mu}\left(y\right)$
are\emph{ not} dynamical variables and remain ``frozen'' like external
sources or \emph{background} fields. One could promote them to four
Stückelberg scalar fields similarly to how it is done for the unimodular
gravity in \citep{Kuchar:1991xd}, but we refrain from doing this.
Further we can recall that the measure transforms as 
\begin{equation}
d^{4}X\sqrt{-g\left(X\right)}=d^{4}y\left|\frac{\partial X}{\partial y}\right|\sqrt{-g\left(X\right)}=d^{4}y\sqrt{-g\left(y\right)}\,,\label{eq:transformation_measure}
\end{equation}
so that 
\begin{equation}
\left|\frac{\partial X}{\partial y}\right|=\frac{\sqrt{-g\left(y\right)}}{\sqrt{-g\left(X\right)}}\,.\label{eq:Jacobian_ratio}
\end{equation}
In this case the action (\ref{eq:unimatter}) assumes the \emph{generally
covariant} form 
\begin{equation}
S_{X}\left[g,\beta,\Phi_{m}\right]=\int d^{4}y\sqrt{-g}\,\beta\left(\mathscr{L}_{m}-\frac{1}{\sqrt{-g\left(X\right)}}\right)\,,\label{eq:unimatter_general_covariant}
\end{equation}
because with respect to further coordinate transformations $y\rightarrow y\left(y'\right)$
the \emph{background} field $\sqrt{-g\left(X\right)}$ transforms
as a \emph{scalar}. Now we can consider an infinitesimal change of
coordinates 
\begin{equation}
y'^{\mu}=y^{\mu}-\xi^{\mu}\left(y\right)\,,\label{eq:shift}
\end{equation}
which leads to the variations given by the Lie derivatives 
\begin{equation}
\pounds_{\xi}\left(-g\left(X\right)\right)^{-1/2}=\xi^{\mu}\partial_{\mu}\left(-g\left(X\right)\right)^{-1/2}\,,\label{eq:Li_back}
\end{equation}
\begin{equation}
\pounds_{\xi}g^{\mu\nu}=-\xi^{\mu;\nu}-\xi^{\nu;\mu}\,,\label{eq:Li_metric}
\end{equation}
\begin{equation}
\pounds_{\xi}\left(\sqrt{-g}\mathscr{L}_{m}\right)=-\frac{\sqrt{-g}}{2}T_{\mu\nu}^{\left(m\right)}\left(\xi^{\mu;\nu}+\xi^{\nu;\mu}\right)\,,\label{eq:Li_Lagrangian_density}
\end{equation}
\begin{equation}
\pounds_{\xi}\sqrt{-g}=\frac{1}{2}\sqrt{-g}g_{\mu\nu}\left(\xi^{\mu;\nu}+\xi^{\nu;\mu}\right)\,,\label{eq:Li_sqrt_g}
\end{equation}
where $\left(\:\right)_{;\mu}=\nabla_{\mu}\left(\:\right)$. We do
not need variations of $\beta$ and of $\Phi_{m}$, as we assume that
their equations of motion hold and result in vanishing variations
of the action. Moreover, at this step we can use equation of motion
for $\beta$: $\left(-g\left(X\right)\right)^{-1/2}=\mathscr{L}_{m}$.
With respect to the diffeomorphisms (\ref{eq:shift}) the action remains
invariant so that 
\[
\int d^{4}y\sqrt{-g}\beta\left[-T_{\mu\nu}^{\left(m\right)}\xi^{\mu;\nu}-\mathscr{L}_{m}g_{\mu\nu}\xi^{\mu;\nu}-\xi^{\mu}\partial_{\mu}\mathscr{L}_{m}\right]=0\,.
\]
Now we can integrate by parts the first two terms, what, under the
assumption that $\xi^{\mu}$ vanishes on the boundary, provides
\[
\int d^{4}y\sqrt{-g}\left[\nabla^{\nu}\left(\beta T_{\mu\nu}^{\left(m\right)}\right)+\partial_{\mu}\left(\beta\mathscr{L}_{m}\right)-\beta\partial_{\mu}\mathscr{L}_{m}\right]\xi^{\mu}=0\,.
\]
Taking into account arbitrariness of $\xi^{\mu}$ this expression
yields 
\begin{equation}
\nabla^{\nu}\left(\beta T_{\mu\nu}^{\left(m\right)}\right)=-\mathscr{L}_{m}\partial_{\mu}\beta\,,\label{eq:General_EMT_nonconserved}
\end{equation}
which should hold on all equations of motion. Hence, we again obtained
last equality in (\ref{eq:div_Einstein_beta}) which now holds for
generic matter with metric EMT. 

Here we should mention that one could substitute the ``unimatter''
constraint (\ref{eq:unimatter_constrain}) directly into the Einstein-Hilbert
action, similarly to what is sometimes done with the unimodular constraint.
However, contrary to the latter case this does not enforces a polynomial
form of this action in the metric. 

\section{Einstein Frame }

As we have seen in the variational formulation ``Einstein transversality''
condition was equivalent to the ``Energy transversality'' condition.
In the integral formulation of the ``Einstein transversality'' we
assumed that the matter is minimally coupled. In these actions (\ref{eq:Vector_for_G})
and (\ref{eq:Chern_Simons_for_G}) one can make a Weyl transformation
\begin{equation}
g_{\mu\nu}=e^{2\omega}\gamma_{\mu\nu}\,,\label{eq:Conformal_trans}
\end{equation}
in order to pass to an Einstein frame. The Weyl transformation above
will not change the form of the axionic term $\alpha\mathcal{F}_{\mu\nu}\mathcal{F}^{\star\mu\nu}\sqrt{-g}$,
while, after a field redefinition (Weyl transformation) of $C^{\mu}$,
the term $\alpha\nabla_{\mu}C^{\mu}\sqrt{-g}$ will preserve its form
as well. However, as it is well known 
\begin{equation}
R\left(g\right)=e^{-2\omega}\left(R\left(\gamma\right)-6\Box_{\gamma}\omega-6\gamma^{\mu\nu}\partial_{\mu}\omega\partial_{\nu}\omega\right)\,,\label{eq:Conformal_Ricci}
\end{equation}
where $\Box_{\gamma}=\gamma^{\mu\nu}\widetilde{\nabla}_{\mu}\widetilde{\nabla}_{\nu}$
and $\widetilde{\nabla}_{\mu}$ is the Levi-Civita connection compatible
with the transformed metric 
\begin{equation}
\widetilde{\nabla}_{\mu}\gamma_{\alpha\beta}=0\text{ .}\label{eq:metric_compatibility_physical}
\end{equation}
 Further, by taking 
\begin{equation}
\omega=-\frac{1}{2}\log\alpha\,,\label{eq:Omega_of_alpha}
\end{equation}
and introducing 
\begin{equation}
\chi=\sqrt{6}\omega=-\sqrt{\frac{3}{2}}\log\alpha\,,\label{eq:Conical_Norm}
\end{equation}
one obtains the action (\ref{eq:Chern_Simons_for_G}) in the Einstein
frame 
\begin{align}
 & S\left[\gamma,\mathcal{A},\chi\right]=\int d^{4}x\sqrt{-\gamma}\left[-\frac{1}{2}R\left(\gamma\right)+\frac{1}{2}\gamma^{\mu\nu}\partial_{\mu}\chi\partial_{\nu}\chi+\right.\nonumber \\
 & \qquad\qquad\left.+e^{-\sqrt{2/3}\chi}\mathcal{F}_{\mu\nu}\mathcal{F}^{\star\mu\nu}\right]\,.\label{eq:Einstein_frame_gauge}
\end{align}
The usual matter is now coupled to the metric 
\begin{equation}
g_{\mu\nu}=e^{2\omega}\gamma_{\mu\nu}=e^{\sqrt{2/3}\chi}\gamma_{\mu\nu}\,.\label{eq:matter_coupling_metric}
\end{equation}
Here we omitted boundary terms. Interestingly, the field $\chi$ which
is just a mere field redefinition of the Lagrange multiplier $\alpha$,
looks now as a healthy scalar field with a usual non-ghosty canonical
kinetic term, though this term does not play any role as it is zero
due to the vector equations of motion, similarly to what happens in
the action (\ref{eq:improvement_by_kinetic}). 

Clearly, for the vector field action one would get 
\begin{align}
 & S\left[\gamma,\widetilde{C},\chi\right]=\int d^{4}x\sqrt{-\gamma}\left[-\frac{1}{2}R\left(\gamma\right)+\frac{1}{2}\gamma^{\mu\nu}\partial_{\mu}\chi\partial_{\nu}\chi+\right.\nonumber \\
 & \qquad\qquad\left.+e^{-\sqrt{2/3}\chi}\widetilde{\nabla}_{\mu}\widetilde{C}^{\mu}\right]\,,\label{eq:Einstein_frame_vector}
\end{align}
where the vector $C^{\mu}$ got transformed according to it's conformal
weight 4: 
\begin{equation}
\widetilde{C}^{\mu}=e^{4\omega}C^{\mu}=\alpha^{-2}C^{\mu}\,.\label{eq:transformation_C}
\end{equation}
Of course we could omit the kinetic term for $\chi$ in these Einstein
frame actions. Then these actions deviate from the (\ref{eq:rescaled_matter_action})
and (\ref{eq:rescaled_matter_axion}) only by matter fields rescaling.
The rescaling of the matter metric is in accord with what was discussed
in the vacuum sequester \citep{Kaloper:2013zca,Kaloper:2014dqa,Kaloper:2015jra,Kaloper:2016jsd,Kaloper:2018kma,Ben-Dayan:2015nva}. 

\section*{One constant of integration to fix both $G_{N}$ and $\Lambda$ }

One can further extend (\ref{eq:axion_for_Newton}) by writing 
\begin{align}
 & S\left[g,\mathcal{A},\varphi\right]=\int d^{4}x\sqrt{-g}\left[-\frac{1}{2}\mu^{2}\left(\varphi\right)R+\alpha\left(\varphi\right)\mathcal{F}_{\mu\nu}\mathcal{F}^{\star\mu\nu}+\right.\nonumber \\
 & \left.\qquad\qquad\qquad\qquad\qquad+\frac{1}{2}\left(\partial\varphi\right)^{2}-V\left(\varphi\right)\right]\label{eq:one_constant_for_both}
\end{align}
where $\mu\left(\varphi\right)$ , $\alpha\left(\varphi\right)$ and
$V\left(\varphi\right)$ are arbitrary functions of the scalar field
$\varphi$ and the normal matter is minimally coupled. One should
assume that $\varphi$ is a pseudoscalar so that $\alpha\left(-\varphi\right)=-\alpha\left(\varphi\right)$.
The gauge field equation of motion (\ref{eq:equation_gauge_field-1})
gets an irrelevant modification 
\begin{equation}
\frac{1}{\sqrt{-g}}\cdot\frac{\delta S}{\delta\mathcal{A}_{\gamma}}=4\alpha'\mathcal{F}^{\star\gamma\mu}\partial_{\mu}\varphi=0\,,\label{eq:constancy_phi}
\end{equation}
where prime denotes derivative with respect to $\varphi$. This equation
has an obvious solution 
\begin{equation}
\varphi=\varphi_{\star}=\text{const}\,.\label{eq:phi_star}
\end{equation}
While the variation with respect to $\varphi$ yields constraint 
\begin{equation}
\alpha'\mathcal{F}_{\mu\nu}\mathcal{F}^{\star\mu\nu}-V'=\mu\mu'R\,.\label{eq:shifted_constraint}
\end{equation}
where we used (\ref{eq:phi_star}). This constraint allows one to
express the Pontryagin invariant, $\mathcal{F}_{\mu\nu}\mathcal{F}^{\star\mu\nu}$,
through the Ricci scalar. The later is to obtain from the tensor equations
of motion. The functional dependence $\mu\left(\varphi\right)$, $\alpha\left(\varphi\right)$
reduces to a mere coefficient of proportionality while the ``potential'',
$V\left(\varphi\right)$ only provides a linear shift, as all these
functions are evaluated at $\varphi_{\star}$. This shift is important
because it allows to have a nonvanishing Pontryagin invariant even
for a vanishing Ricci scalar. 

In this Jourdan frame one obtains that the effective Planck mass is
given by 
\begin{equation}
M_{Pl}=\frac{\mu\left(\varphi_{\star}\right)}{\sqrt{8\pi}}\,,\label{eq:Planck_mass}
\end{equation}
while the potential provides the effective cosmological constant 
\begin{equation}
\Lambda=V\left(\varphi_{\star}\right)\,.\label{eq:V_Lambda}
\end{equation}
Thus one integration constant provides the values for both coupling
constants of GR. Let us briefly discuss the functions which can appear. 

Simplest option from dimensional reasons would be $\mu\left(\varphi\right)=\varphi.$
If the gauge filed is abelian, it can have arbitrary dimensionality
as we neglect the standard kinetic term anyway. In particular, $\left[\mathcal{A}\right]=\left[m^{1/2}\right]$
would allow for the simplest coupling $\alpha=\varphi.$ On the other
hand, the form of the covariant derivative $D_{\mu}=\partial_{\mu}+i\text{g}\mathcal{A}_{\mu}$
with the self-coupling constant $\text{g}$ dictates that either the
coupling constant $\text{g}$ is dimensionless and $\left[\mathcal{A}\right]=\left[m\right]$
so that one has to introduce a scale $M$ into the axion coupling
$\alpha=\varphi/M$, or that $\left[\mathcal{A}\right]=\left[m^{1/2}\right]$
and $\left[\text{g}\right]=\left[m^{1/2}\right]$. Thus the non-abelian
realization provides a scale which can be used in the potential. In
the later case one could write the potential as 
\begin{equation}
V\left(\varphi\right)=\text{g}^{8}f\left(\frac{\varphi}{\text{g}^{2}}\right)\,.\label{eq:general_potential}
\end{equation}
Appealing forms of the potential would resemble instanton contribution 

\begin{equation}
V_{1}\left(\varphi\right)=\frac{1}{4}\varphi^{4}\exp\left(-\frac{\varphi}{\text{g}^{2}}\right)\,,\qquad\label{eq:inst_1}
\end{equation}
and 
\begin{equation}
V_{2}\left(\varphi\right)=\frac{1}{4}\text{g}^{8}\exp\left(-\frac{\varphi}{\text{g}^{2}}\right)\,.\label{eq:inst_2}
\end{equation}
It would be great if the appearance of the scale in the coupling constant
$\text{g}$ could be related to the scale of strong coupling or dimensional
transmutation. The strong coupling may hint to an explanation of the
absence of the usual kinetic term for the gauge field in the action
(\ref{eq:one_constant_for_both}). To obtain 120 orders of magnitude
difference one would need to assume: for $V_{1}\left(\varphi\right)$
that $\text{g}^{2}\simeq\varphi_{\star}/277$, whereas for $V_{2}\left(\varphi\right)$
that $\text{g}^{2}\simeq\varphi_{\star}/254$. 

Finally we would like to note that the sign of the kinetic term for
$\varphi$ can also be negative. Moreover, one can write an arbitrary
function $P\left(\left(\partial\varphi\right)^{2},\varphi\right)$.
Similarly to \citep{Lim:2010yk} the functional dependence on kinetic
term does not change anything here. 

\section*{Conclusions }

In this paper we proposed a new very simple scale-free Einstein equations
(\ref{eq:normalized-2}) which promote the Newton constant to a global
degree of freedom. In this way the trace part of the usual Einstein
equations is exchanged to a useless identity $1=1$. This is similar
to trace--free Einstein equations of the unimodular gravity where
this identity is $0=0$. Then we discussed different equivalent formulations
of these equations and their variational formulations. After that
we introduced different generally covariant actions (\ref{eq:Vector_for_G}),
(\ref{eq:Chern_Simons_for_G}), (\ref{eq:rescaled_matter_action}),
(\ref{eq:rescaled_matter_axion}), (\ref{eq:improvement_by_kinetic}),
(\ref{eq:one_constant_for_both}) describing Newton constant as a
global degree of freedom. From these actions, we derived and discussed
novel non-covariant ``unicurvature'' (\ref{eq:unicurvature_action})
and ``unimatter'' (\ref{eq:unimatter}) actions following the same
ideology as many works on unimodular gravity. It is interesting whether
such non-covariant formulations can be further meaningfully generalized,
as it was done in the case of unimodular gravity in \citep{Barvinsky:2017pmm,Barvinsky:2019agh,Barvinsky:2020sxl}.
Some of our frameworks for the dynamical Newton constant (\ref{eq:rescaled_matter_action}),
(\ref{eq:rescaled_matter_axion}), (\ref{eq:improvement_by_kinetic})
actually imply that the effective Planck constant $\hbar$ becomes
a global degree of freedom, similarly to claims of \citep{Carroll:2017gqo}
where a different system was discussed. We have also written the Heisenberg
uncertainty relations (\ref{eq:uncertainty_G_alpha}), (\ref{eq:matter_uncertainty_Realtion})
for the effective Newton constant $G_{N}$; (\ref{eq:uncertainty_volume})
for cosmological constant $\Lambda$; (\ref{eq:uncertainty_for_Planck})
for the Planck constant $\hbar$ and (\ref{eq:uncertainty_Planck_length})
for the Planck length $\ell_{Pl}$. Most probably these relations
can only be physically relevant close to singularities either at the
very beginning of our universe or at the final stages of gravitational
collapse inside black holes. We left any detailed discussion of these
uncertainty relations for future work. 

Intriguingly, the unimodular gravity naturally appears in the thermodynamic
or emergent gravity setup \citep{Jacobson:1995ab,Padmanabhan:2013nxa,Padmanabhan:2019art,Alonso-Serrano:2020dcz}.
There are already some hints \citep{Volovik:2020qtp,Klinkhamer:2008ff}
that the scale-free gravity (\ref{eq:normalized-2}) with the Newton
constant as a constant of integration, may also emerge in this way.
It is very interesting to further investigate this option. 

On the other hand, combining unimodular gravity with theories where
the Newton constant is another global degree of freedom is the key
element of the vacuum sequester proposal, \citep{Kaloper:2013zca,Kaloper:2015jra,Padilla:2015aaa}.
A novelty with respect to vacuum sequester works \citep{Kaloper:2013zca,Kaloper:2014dqa,Kaloper:2015jra,Kaloper:2016jsd,Kaloper:2018kma,Ben-Dayan:2015nva,Carroll:2017gqo}
is that following \citep{Hammer:2020dqp} we also use normal Yang-Mills
gauge fields instead of more exotic three- and four-forms to specify
a new measure for the action integral. Moreover, in our formulation
the Lagrange multiplier fields from these setups look like axions
and can even have normal kinetic terms, see actions (\ref{eq:axion_for_Lambda}),
(\ref{eq:axion_for_Newton}) and (\ref{eq:improvement_by_kinetic}).
For the Newton constant to be a global degree of freedom the axion
should either non-minimally couple to curvature as in (\ref{eq:axion_for_Newton})
or to couple universally to all matter fields, as in (\ref{eq:improvement_by_kinetic}).
To describe sequester one requires to have at least two different
gauge fields and two different axions. It is rather intriguing to
understand quantum properties of such systems and their potential
UV completions. In particular, such formulation can be also useful
for supersymmetric generalizations, see e.g. \citep{Nagy:2019ywi,Baulieu:2020obv,Bansal:2020krz}.
Even though QCD seems to possess a composite three-form \citep{Luscher:1978rn,Aurilia:1980xj,Dvali:2005an,Dvali:2005ws,Dvali:2005zk},
we think it is more convenient to see usual Yang-Mills gauge fields
in the action. This brings the field content of such theories closer
to elementary fields present already in the Standard Model. This may
be a hint that it is the IR dynamics of the Yang-Mills vacuum (maybe
even QCD or some ``dark'' QCD) and its topological properties which
secretly define both gravitational constants $G_{N}$ and $\Lambda$. 
\begin{acknowledgments}
It is a pleasure to thank I.\,Saltas and I.\,Sawicki, for valuable
discussions and criticisms, while we are indebted to N.\,Kaloper
and T.\,Padilla for very useful email exchange. K.\,S. would like
to thank the Central European Institute for Cosmology and Fundamental
Physics for their warm and welcoming hospitality during early stages
of the project. K.\,S. is supported by JSPS KAKENHI Grant Number
JP20J12585. M.\,Y. is supported in part by JSPS Grant-in-Aid for
Scientific Research Numbers 18K18764 and JSPS Bilateral Open Partnership
Joint Research Projects. This project originated when P.\,J. and
A.\,V. were enjoying very warm hospitality of the cosmology group
at the Tokyo Institute of Technology. This productive visit was possible
thanks to the JSPS Invitational Fellowships for Research in Japan
(Fellowship ID:S19062) received by A.\,V. The work of P.\,J. is
supported by the Grant Agency of the Czech Republic GA\foreignlanguage{american}{\v{C}}R
grant 20-28525S. The work of A.\,V. is supported by the J.\,E.\,Purkyn\foreignlanguage{american}{\v{e}}
Fellowship of the Czech Academy of Sciences and by the funds from
the European Regional Development Fund and the Czech Ministry of Education,
Youth and Sports (M\foreignlanguage{american}{\v{S}}MT): Project
CoGraDS - CZ.02.1.01/0.0/0.0/15\_003/0000437. 
\end{acknowledgments}

\bibliographystyle{utphys}
\phantomsection\addcontentsline{toc}{section}{\refname}\bibliography{Unimod}

\end{document}